\documentclass{iaus}
\usepackage{graphicx}
\usepackage{natbib}		

\DeclareTextSymbol{\degre}{T1}{6}
\DeclareTextSymbol{\degre}{OT1}{23}

\newcommand{\aaps}{A\&AS}

\newcommand{\mnras}{MNRAS}

\title[The constant magnetic field of $\xi^1$~CMa.] 
{The constant magnetic field of $\xi^1$~CMa : geometry or slow rotation ?}

\author[C. Fourtune-Ravard and the MiMeS Collaboration]   
{C. Fourtune-Ravard$^{1,2}$, G.A. Wade$^2$, W. Marcolino$^{3,4}$, M. Shultz$^5$, J. Grunhut$^{2,5}$, H. Henrichs$^6$ and the MiMeS Collaboration}

\affiliation{
$^1$Universit\'e Paris Diderot-Paris 7, UFR de Physique, France
 \\ email: {\tt c.fourtune.ravard@gmail.com} \\[\affilskip]
$^2$Department of Physics, Royal Military College of Canada, Ontario, Canada\\[\affilskip]
$^3$LAM-UMR 6110, CNRS \& Univ. de Provence, France \\[\affilskip]
$^4$Observat\`orio Nacional, Rio de Janeiro, Brazil \\[\affilskip]
$^5$Department of Physics, Engineering Physics \& Astronomy, Queen's University, Canada \\[\affilskip]
$^6$Astronomical Institute ‘Anton Pannekoek’, University of Amsterdam, the Netherlands
}

\pubyear{2010}
\volume{272}  
\pagerange{119--126}
\jname{Active OB stars: structure, evolution, mass loss and critical limits}
\editors{A.C. Editor, B.D. Editor \& C.E. Editor, eds.}
\begin{document}

\maketitle

\begin{abstract}
We report recent observations of the sharp-lined magnetic $\beta$~Cep pulsator $\xi^1$~CMa (= HD 46328). The longitudinal magnetic field of this star is detected consistently, but it is not observed to vary strongly, during nearly 5 years of observation. In this paper we evaluate whether the nearly constant longitudinal field is due to intrinsically slow rotation, or rather if the stellar or magnetic geometry is responsible.
\end{abstract}

\section{Introduction, observations and stellar parameter determination}

$\xi^1$~CMa is known to be an B0.5 pulsator with sharp lines. The longitudinal magnetic field of this star, first detected by \cite{hubrig06} with FORS1, is detected consistently in over 5 years of observations, but has remain approximately constant at Bz $\sim$375 G. Within the rigid rotator paradigm, two explanations can explain this behaviour: either the star rotates very slowly, or the stellar or magnetic geometry is responsible for the constant value of the longitudinal magnetic field. We acquired 18 Stokes $V$ ESPaDOnS spectra ($370\leq\lambda\leq 1000$~nm, $R=65,000$, S/N$\sim 1000$ per 1.8 km/s pixel) with the aim of precisely studying the magnetic field, investigating the rotational period and geometry.


We employed CMFGEN to determine physical and wind parameters of $\xi^1$~CMa. The luminosity was computed using the parallax of \cite{leeuwen07}, which provides a good fit to the IUE (SWP+LWR) low resolution and large aperture data. An E(B-V) of 0.04 was also used. We find $T_{\rm eff}=27500 \pm 2000$~K, $\log g=3.50 \pm 0.20$, $L/L_\odot=38370$, $R/R_\odot=8.6$ and $M/M_\odot\sim 9.0$. The projected rotational velocity is constrained to be $v\sin i\leq 15$~km/s.

\section{Magnetic field and rotational period}
Longitudinal magnetic field measurements were inferred using Least-Squares Deconvolution (LSD) with a line mask carefully customised to the spectrum of $\xi^1$~CMa. All spectra yield definite detections of Stokes $V$ profiles and flat diagnostic N profiles with longitudinal field uncertainties of $\sim$7 G. A straight-line fit to the longitudinal field measurements extracted from Stokes $V$ gives a reduced $\chi ^2$ of 4.8,  while analogous measurements extracted from diagnostic $N$ give reduced $\chi ^2$ of just 1.1. This points to weak variability of Stokes $V$ that is not present in $N$.

$\xi^1$~CMa is a well-known $\beta$~Cep pulsator that displays monoperiodic radial mode photometric and line profile variability with a period of 0.209 days (\cite{heynderickx94}; \cite{saesen06}). Our spectra sample the full pulsational cycle and reveal a peak-to-peak radial velocity variation of 38 km.s$^{-1}$. We perfomed a period search of the Stokes $V$ longitudinal field measurements using a Lomb-Scargle algorithm, detecting significant power at 4.2680 days. When the longitudinal field measurements are phased with this period they describe a sinusoidal variation with amplitude $\sim$30 G and reduced  $\chi ^2$ of 1.2. 

\section{H$\alpha$ emission and UV wind line morphology}
We observed emission in the H$\alpha$ line profile. We extracted the emission profile from each spectrum by first using H$\beta$ to construct a photospheric template, then subtracting a model photospheric profile from the H$\alpha$ profile. The derived emission profile is approximately constant and characterised by a FWHM of $\sim$120 km.s$^{-1}$. The UV C~{\sc iv} and Si~{\sc iv} wind lines of $\xi^1$~CMa show no variability in IUE spectra acquired in 1978 and 1979. They are remarkably similar to those of the magnetic star $\beta$~Cep at phases of maximum emission (i.e. when the star is viewed closest to the magnetic pole). This could imply that we currently view $\xi^1$~CMa near its magnetic pole as well. Such a configuration is potentially consistent with either a long rotation period or a pole-on geometry. Nevertheless, the lack of variability observed in the UV and optical wind lines leads us to prefer the pole-on geometry model. 

\section{Conclusions: magnetic field, stellar geometry and rotation}
The lack of any secular change in the field during the period of observation, in combination with the very high precision of the magnetic measurements, suggests either that the rotational period of the star is remarkably long, or that the stellar geometry is such that the disc-integrated line-of-sight component of the field remains approximately constant. The latter model is more consistent with the observed stability of the H$\alpha$ and UV line emission, the UV line morphology, and the period detected in the longitudinal field measurements. If $\beta$~Cep has $i$=60$\degre$ and $\beta$=85$\degre$, the magnetic pole comes within 35$\degre$ of the line of sight. If we accept the arguments above, the magnetic pole of $\xi^1$~CMa must come similarly close to the line of sight, and furthermore it must remain there. This would all hold together if the 4.26 day period is in fact the stellar rotation period. This would require very low inclination (5 to 10$\degre$) to be consistent with the low vsini, and in that case the weak modulation of the longitudinal field can be matched by a 1450 G dipole with obliquity $\beta$=25$\degre$. This would imply, in fact, that the magnetic pole is never more than 30$\degre$ from the line of sight. An accurate determination of the $v\sin i$ of this star is critical to confirming this view, as effectively any non-zero value of $v\sin i$ would imply that the rotational period is shorter than the total span of the observations.

\end{document}